# HZO-based FerroNEMS MAC for In-Memory Computing

Shubham Jadhav[1], Ved Gund[1], Benyamin Davaji[3], Debdeep Jena[1,2], Huili (Grace) Xing[1,2], and Amit Lal[1]

[1]School of Electrical and Computer Engineering, Cornell University, Ithaca, NY, USA email: saj96@cornell.edu
[2]Department of Materials Science and Engineering, Cornell University , Ithaca, NY, USA
[3]Electrical and Computer Engineering department, Northeastern University, Boston, MA, USA

**Abstract:** This paper demonstrates a hafnium zirconium oxide (HZO)-based ferroelectric NEMS unimorph as the fundamental building block for very low-energy capacitive readout in-memory computing. The reported device consists of a 250 $\mu m \times$ 30 $\mu m$ unimorph cantilever with 20 nm thick ferroelectric HZO on 1 µm SiO$_2$. Partial ferroelectric switching in HZO achieves analog programmable control of the piezoelectric coefficient (d$_{31}$) which serves as the computational weight for multiply-accumulate (MAC) operations. The displacement of the piezoelectric unimorph was recorded by actuating the device with different input voltages V$_{in}$. The resulting displacement was measured as a function of the ferroelectric programming/poling voltage $V_P$. The slopes of central beam displacement (δ$_{max}$) vs V$_{in}$ were measured to be between 182.9nm/V (for –8V$_p$) and –90.5nm/V (for 8V$_p$), demonstrating that V$_p$ can be used to change the direction of motion of the beam. The resultant δ$_{max}$ from AC actuation is in the range of –18 to 36 nm and is a scaled product of the input voltage and programmed d$_{31}$ (governed by the V$_p$). The multiplication function serves as the fundamental unit for MAC operations with the ferroelectric NEMS unimorph. The displacement from many such beams can be added by summing the capacitance changes, providing a pathway to implement a multi-input and multi-weight neuron. A scaling and fabrication analysis suggests that this device can be CMOS compatible, achieving high in-memory computational throughput.

Neuromorphic computation is of great interest to computing theory and practical implementations due to the potential for low-power, high efficiency, and small form factor information processing with deep neural networks (DNNs) [1]–[3]. The building blocks of DNNs are perceptron blocks, which inherently necessitate brain-like synaptic functions achieved through arrays of MAC units. With the ever-increasing number of variables required for neuromorphic computation with high accuracy, there is an urgent need to develop highly energy-efficient device architectures [3]–[5]. The energy used per MAC unit is, therefore, a useful metric to compare technologies. Ferroelectric field-effect transistors (Fe-FET) have been used to mimic analog synaptic operation with gate-ferroelectric switching in silicon and germanium channel devices to realize multi-level conductance modulation with potentiation and depression with 50-100 ns switching pulses and 3-5 MV/cm fields[6], [7]. FET-based in-memory computation architectures are susceptible to large read and write energy consumption and high leakage currents in idle mode,  particularly with gate dielectric thickness scaling down to < 5 nm [8]. NEMS switches and beams offer an alternate pathway to zero-leakage in-memory compute synaptic functionality, provided that the beam actuation has embedded programmable weights in the form of tunable capacitive or piezoelectric coupling. We have previously reported a graphene-on-HZO ferroelectric device that uses potential gradients across the graphene to achieve fractional switching, with the modulated graphene resistance also used for resistive memory readout [9]**.** While analog in-memory computing has been demonstrated using different architectures which use transistors or memristors, few prior works have used a NEMS-based approach that takes advantage of a released beam structure to eliminate energy leakage in an idle state [10]–[13]. In previous work, the idea of using nanomechanical electrostatic actuators to produce MAC units has been described [13]. In contrast, we present a ferroelectric/piezoelectric beam transducer to enable the multiplication that can be read out capacitively, eliminating any DC currents.

This paper presents a novel technique to store the neural net weights in the form of the programmable piezoelectric coefficient $d_{31}$ of HZO-driven unimorph, which changes by applying a polarization switching voltage $V_P$. After polarization programming, different actuation inputs ($V_{in}$), actuate the cantilever such that the unimorph displacement is a scaled product of $V_p$ and $V_{in}$. The displacement can in turn be measured as capacitive sense current from a capacitive divider circuit, for massively parallel MAC operations from arrays of such NEMS elements. Figure 1 shows the concept of the partial polarization in a ferroelectric for analog control of beam-bending in a unimorph. In addition to the active ferroelectric film, the unimorph stack also includes metal contact layers and an insulating elastic layer. The polarization versus E-field (PE loop) for an 80 μm diameter metal-ferroelectric-metal (MFM) capacitor on a 20 nm thin HZO film is shown in figure 1a. Moving counterclockwise along the PE loop starting with all dipoles point downward (point O), the film can be in a state of net-zero polarization (points A and A') at the positive and negative coercive fields, fully down or up-switched (points C and C') or partially polarized (along the slopes of A-B and A'-B'). Each point along the PE loop achieves a specific macroscopic value of $d_{31}$, with the maximum and minimum values at the two extrema (at points C and C') with the appropriate choice of bias voltage $V_p = E t_{HZO}$. After programming, if a much smaller actuation voltage $V_{in}$ is applied such that it produces no dipole switching, an in-plane stress is generated due to expansion or contraction of the HZO film along its length, resulting in the piezoelectric bending moment $M_{Piezo}$. The programming would enable weight-storage in DNNs at reduced rates compared to inference events. The beam displacement is a solution of the Euler–Bernoulli equation:

$$\frac{\partial^2 y}{\partial x^2} = -\frac{M_{Piezo}}{C} \tag{1}$$

Here, $M_{piezo}$ is the piezoelectric moment generated due to applied $V_{in}$ and $C$ is the total flexural rigidity of the multi-layer piezoelectric stack [14]. By using zero displacements and zero slopes boundary conditions on each end of the clamped-clamped beam, the Euler–Bernoulli equation yields the beam displacement profile. The maximum beam displacement occurs at the center i.e. $y(\frac{1}{2}) = \delta_{max}$. For piezoelectric coefficient $d_{31}$ and in-plane stress $\sigma_1$ in HZO, the beam displacement can be expressed as:

$$\delta_{max} \propto M_{Piezo} \propto \sigma_1 \propto d_{31} \propto P(Polarization) \propto V_p \tag{2}$$

For an input voltage $V_{in}$, the NEMS beam displacement is, hence, a scaled product of weights and inputs.

$$\delta_{max} = SF \cdot V_p \cdot V_{in} \tag{3}$$

Here, the scale factor (SF) depends on ferroelectric material properties, beam geometry, Young's modulus, etc. However, we can design the multiplier to achieve linear behavior in the transfer characteristics curve along the rising and falling slopes of the PE loop to accomplish a multiplicative function for parallel MAC operations. Figure 1c represents the multiplier unit with a capacitive readout to measure the output voltage. The left and right beams are poled with the same magnitude but with opposite polarity. Thus, for the same input voltage $V_{in}$, the left and right beams displace such that the difference between the two represents the product of $V_{in}$ and $V_P$. The differential motion causes differential capacitance between left and right capacitive parallel plates. The output voltage is given by:

$$V_{out} = 2V_{dd} \frac{C_L - C_R}{C_L + C_R} \cong -2V_{dd} \frac{\gamma(x_L - x_R)}{2g} = SF \cdot V_p \cdot V_{in} \tag{4}$$

Where $\gamma$ is an constant which depends on beam geometry, and $x_L$ and $x_R$ are left and right parallel plate displacements with initial gap g. Several multiplier units can then be further connected in parallel with an activation unit to form a perceptron. The output voltage of the MAC unit (figure 1d) is:

$$V_{out,MAC} = 2V_{dd} \frac{\sum_{i=1}^{N} C_{L,i} - \sum_{i=1}^{N} C_{R,i}}{\sum_{i=1}^{N} C_{L,i} + \sum_{i=1}^{N} C_{R,i}} = SF \sum_{i=1}^{N} V_{p,i}.V_{in,i} \qquad (5)$$

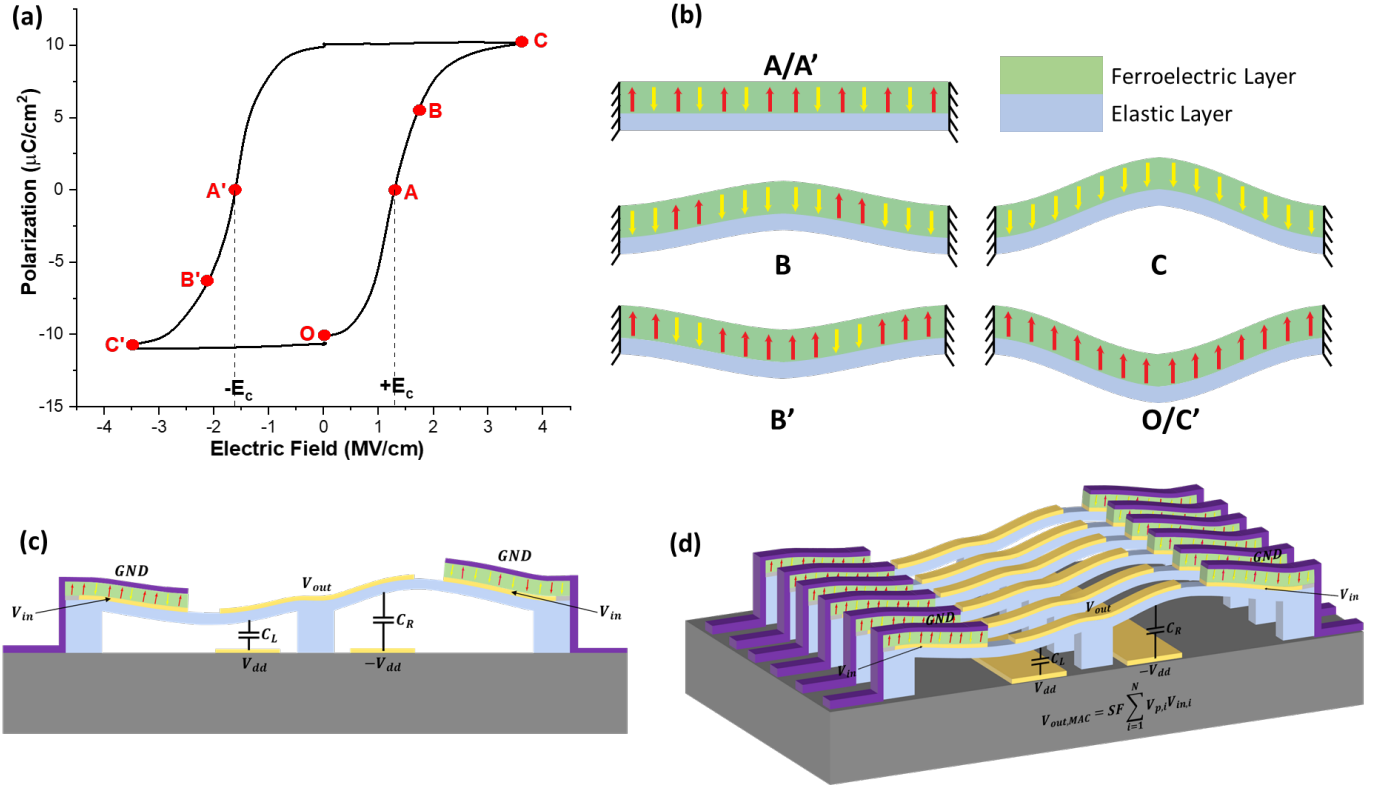

**FIG 1** *(a) PUND loop of 20 nm-thick HZO film (b) Polarization-dependent bending moment generated in ferroelectric-on-elastic unimorph at various points along the PE loop ($E_C$ is coercive field) (c) Conceptual multiplier unit with ferroelectric actuation and capacitive readout. The output voltage is the product of $V_{in}$ and stored weights in the form of $d_{31}$ coefficients. (d) Several multipliers with capacitive readout can be placed in parallel to form a MAC unit that can perform accumulation operation by summing up the capacitances from each device.*

Figure 2a and 2b show the top and cross-section views respectively of the clamped-clamped ferroelectric unimorph used to demonstrate piezoelectric coefficient programming. 1 µm thick thermal $SiO_2$ forms the elastic layer underneath 20 nm ferroelectric HZO. The HZO is capped by 3 nm of alumina ($Al_2O_3$), followed by annealing at 400°C to crystallize the HZO in its ferroelectric orthorhombic phase. 200 nm platinum (Pt) and 100 nm aluminum (Al) were deposited to form the bottom and top metal contacts for the HZO respectively. The beam was released by isotropic etching of the silicon substrate using Xactix $XeF_2$ etcher. Details of the fabrication process and challenges encountered are included in the supplementary material. A scanning electron microscope image of the fabricated device is shown in Figure 2c. After release, the beams were observed to be buckled due to residual film-stress generated during microfabrication. A 3D optical profilometer (Zygo$^{TM}$ system) was used to measure the beam buckling profile (figure 2d). The maximum displacement for a $250 \ \mu m \times 30 \ \mu m$ was 4.98 µm.

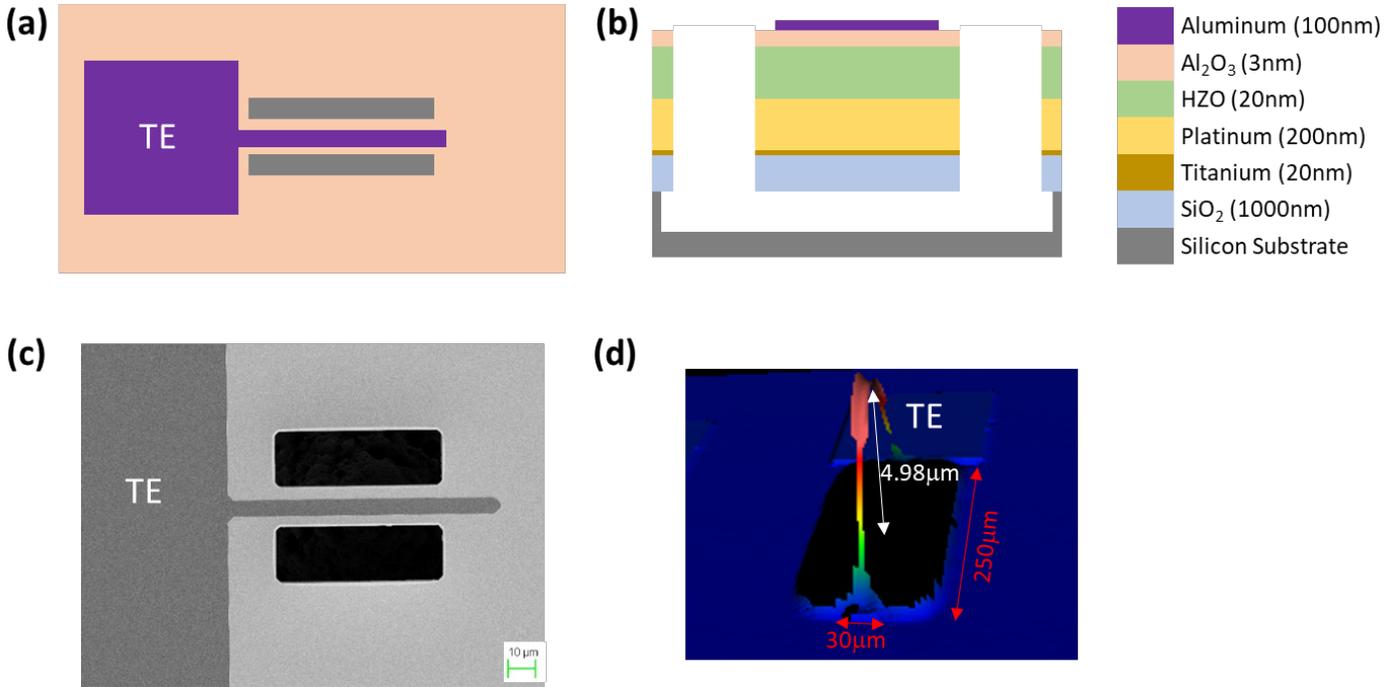

*FIG 2 (a),(b) Schematic top and cross-section view of the ferroelectric beam (c) SEM image of the beam showing released structure (d) 3D profile of the released beam measured using Zygo$^{TM}$ profilometer showing beam curvature due to residual stresses in the stack. For 250× 30μm beam, the measured maximum displacement at the center was 4.98μm.*

HZO ferroelectric characterization was performed using a Sawyer-Tower circuit configuration on a custom probe station with continuous wave (CW) positive-up-negative-down (PUND) input waveforms to extract the coercive field ($E_C$) and remnant polarization ($P_r$). In prior work, we have used electrical breakdown under a large top metal pad to contact the bottom electrode with 5-10Ω resistance [15]. The PUND input signal has 83 μs rise, fall, and wait times (corresponding to 1kHz PUND frequency), and a peak-to-peak amplitude of 15V. Figure 3a shows time-domain input voltage and output sense current measured with an 11.8 kΩ sense resistor. The P and N pulses show a switching current peak below the peak voltage, indicative of polarization switching, whereas the U and D pulses show capacitive displacement current and low leakage. This confirms ferroelectricity in the released HZO films without any degradation, consistent over 10,000 cycles of switching with CW operation. Figure 3b plots the switching current density (J) vs E-field. The PUND measurement was performed on multiple such released NEMS beams with a range of lengths from 100-400 μm and widths spanning 15-39 μm. The overlay from these measurements is shown with multiple dotted lines in figure 3c. For comparison, the PUND loop for an unreleased 80 μm diameter circular electrode is also shown on the same plot (solid black line). This shows that there is little/no degradation of ferroelectric properties of the HZO before and after release. Multiple release devices with different electrode sizes had overlapping loops suggesting high repeatability. PUND measurements yield an extracted $E_c \approx 1.5 MV/cm$ and $P_r \approx 10 \mu C/cm^2$, consistent with previously reported values[16]–[18]. The devices were also tested for DC breakdown as shown in figure 3d. The device's current density exceeds 10μA/cm² at $3.15 MV/cm$.

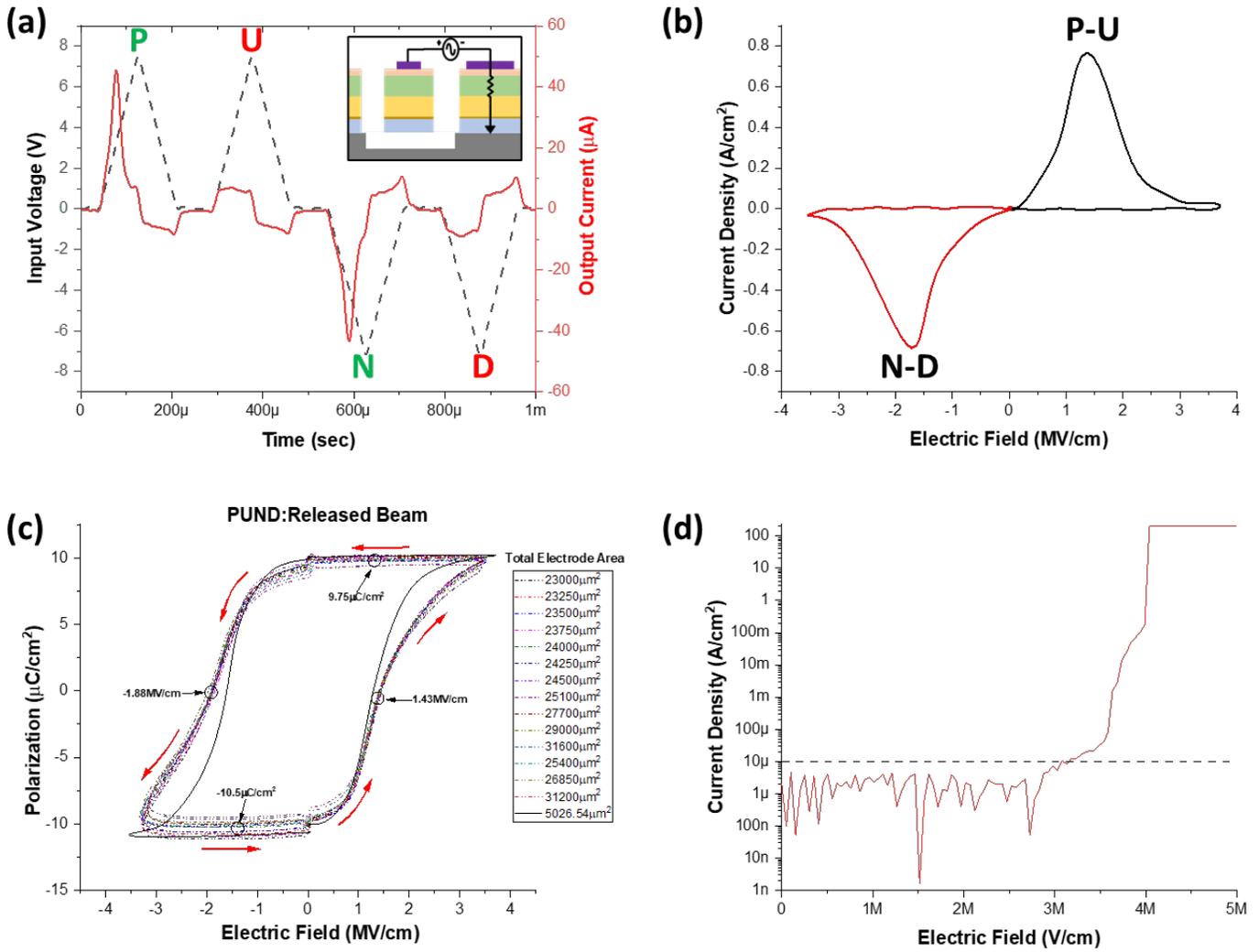

**FIG 3** *(a) Input voltage and output current vs time plot of the device showing much larger switching currents as compared to leakage currents. Inset- Setup for poling the NEMS beam (b) Switching current density vs electric field plot showing ferroelectric switching peaks. (c) Measured Polarization vs E-field characteristics of released beam devices with different lengths and widths showing repeatability. The Measured PE loop of the standard unreleased circular electrode is shown with a solid black (d) DC I-V of the device showing high current density above 3.15 MV/cm indicating leakage.*

The central device concept of this work is to use analog weight storage in the bending motion of the NEMS beam for in-memory computation, which relies on the programming of the $d_{31}$ coefficient of the device by using different poling voltages. The test methodology used for this experiment is as follows:

**Step 1:** Device poling with CW single-sided pulses at 3 kHz. The configuration used for poling is depicted in figure 3a (inset). The test signal is applied to the top electrode, and the ground connection is made to the bottom electrode by breaking down the HZO under a large top metal pad to short it to the bottom electrode.

**Step 2:** Beam actuation with small $V_{in}$ to measure the displacement amplitude, without additional polarization switching.

**Step 3:** Increase poling voltage by 0.5V and repeat steps 1-2, while tracing the PE loop counterclockwise from 0V→8V→0V→-8V→0V.

This voltage sweep for $V_p$ is similar to the PUND loop with one key difference. In this technique, we stepped the $V_p$ by 0.5V and then measured the beam displacement with a fixed $V_{in}$ amplitude, unlike the PUND measurements in CW operation with increasing voltages.

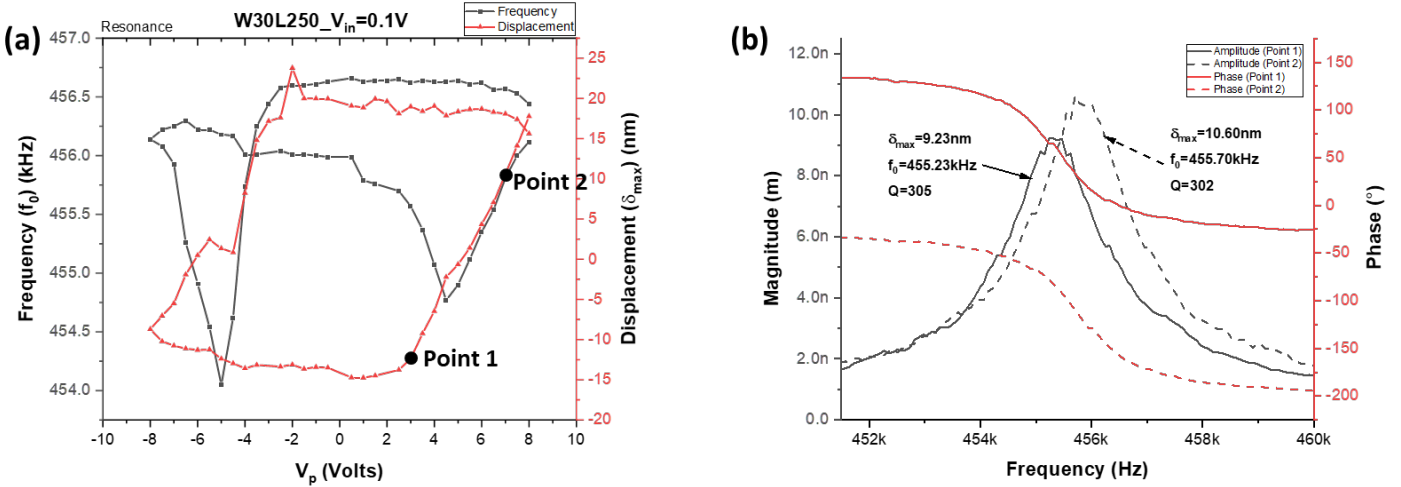

**FIG 4** *(a) Central beam displacement $\delta_{max}$ and resonant frequency $f_0$ vs poling voltage $V_p$ plot for the 250 μm × 30 μm ferroelectric clamped-clamped beam. Devices were poled at different $V_p$ with 0.5V increment along the loop (0V→8V→0V→-8V→0V) and the beam displacement profile was measured using the Polytec LDV at $V_{in}$ =0.1 V in the frequency range of 450-460kHz. The $\delta_{max}$ (red line) changes for different values of $V_p$ and traces a hysteresis loop similar to PUND measurements. The polarization in the beam, induced due to $V_p$, changes the internal stress resulting in the frequency $f_0$ tuning (black line) which could be exploited for memory storage. (b) The resonance vibration magnitude and phase plots of the same beam at different points on plot (a).*

A Polytec MSA-400 laser doppler vibrometer (LDV) was used to measure the beam motion. For a 250 μm × 30 μm beam, a 0.1V amplitude frequency chirp from 1kHz to 1 MHz was applied. A resonance at $f_0$=455.23 kHz was identified with a quality factor of 305. AC voltage $V_{in}$=0.1V was used to measure peak amplitudes. Figure 4a shows the peak beam displacement $\delta_{max}$ and resonant frequency $f_0$ vs poling voltage $V_p$ plot for the nominal 250 μm × 30 μm ferroelectric clamped-clamped beam. $\delta_{max}$ (red line) is modulated for different values of $V_p$ and traces a hysteresis loop, similar to the PE loop from PUND measurements. The net effect of the number of upward and downward pointing dipoles that control the macroscopic polarization in the beam (induced due to $V_p$) also changes the beam stiffness resulting in resonance frequency modulation (black line), which presents a separate modality of weight storage in the beam. The two dips in $f_0$ correspond to the positive and negative coercive fields of the HZO film, where the net polarization is almost zero. As illustrated in figure 1b, domain randomization is a potential cause for likely film relaxation at these points causing $f_0$ to drop.

From figure 4a, the positive linear dependence of $\delta_{max}$ can be seen in the voltage interval 2.5V→8V. Similarly, negative linear dependence in the interval -2.5V→-8V. Here, the slope of $\delta_{max}$ vs. $V_p$ is SF×$V_{in}$ as defined in Eq. 3. The effect of $V_{in}$ on $\delta_{max}$ is explained in the next section. Two points were selected on the positive slope to show the resonance vibration magnitude and phase plots of the vibrating beam (figure 4b). Points 1 and 2 have almost the same displacement magnitude but exactly opposite phases. This 180° phase difference is a signature of polarization inversion corresponding to a sign change for $\delta_{max}$. To show the effect of $V_{in}$ on $\delta_{max}$, $V_{in}$ was swept from 0.02V to 0.20V for different values of the $V_p$ (figure 5).

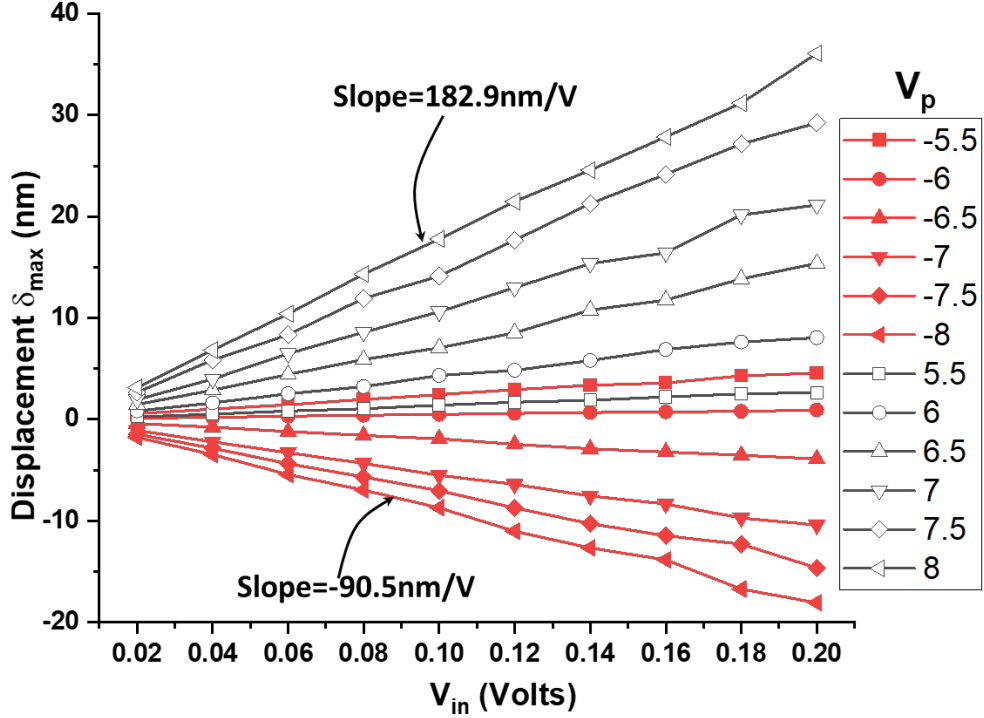

**FIG 5** The displacement vs $V_{in}$ for different values of $V_p$. The black and red traces correspond to the positive $V_p$ (5.5→8V) red negative $V_p$ (-5.5→-8V) poling voltages respectively, showing the multiplication functionality.

Figure 5 plots $\delta_{max}$ vs $V_{in}$ at different values of $V_p$ which emulates the transfer characteristics of the analog multiplier. For a fixed poling voltage, an input voltage sweep yields a linear increase in displacement value. Similarly, for constant inputs, different poling voltages outputs different displacement values. We can calculate the scale factor from previous plots as follows:

$$SF = \frac{\delta_{max}}{V_p V_{in}} \quad (6)$$

Eq. 6 demonstrates that $\delta_{max}$ is the product of inputs ($V_{in}$) and weights ($V_p$). As shown in figure 4a and figure 5, Eq. 6 behaves linearly in the range of $V_p\ from\ \pm 2.5V\ to\ \pm 8V$ and $V_{in}$ from 0 to 0.2V for the device tested here (250 μm×30 μm beam). The beam displacement can be measured using a capacitive divider circuit to measure $V_{out}$. The slope of the plot in figure 5 is the experimental SF×$V_p$. For $V_p = \pm 8V$, the measured slopes were 182.9nm/V and -90.5nm/V respectively. The asymmetry in the negative and positive $V_p$ slopes is likely due to the non-symmetric device stack. Above HZO, there is a very thin top electrode (Al) but below it, we have a thick $SiO_2$ and electrode (Pt) stack. Thus, for the same $V_p$, actuation with opposite polarity will have an unequal effect on the partial polarization resulting in different $\delta_{max}$ values.

In conclusion, we have demonstrated the successful working of an HZO-based NEMS multiplier with weight storage functionality. The device was fabricated and characterized showing the dependence of beam displacement on the poling voltage of the ferroelectric film. Polarization-dependent frequency-tuning was also observed, which can be further explored to achieve an electrically reconfigurable filter. For a $250\ \mu m \times 30\ \mu m$ beam poled at 8V, $\delta_{max}$ of 18 nm and $f_0$ of 455.23 kHz with Q=305 was measured. The $\delta_{max}$ vs $V_{in}$ plot at different values of $V_p$ demonstrate the successful operation of an analog NEMS multiplier. While in this paper AC drive was used to obtain observable displacements, in the future DC input voltages are anticipated for DNN applications. A capacitive readout of the beam motion can be used instead of a topical readout for much higher resolution.

The dimensional and voltage scaling of the prototype device demonstrated is important for CMOS-integrated adoption while maximizing computational throughput. For CMOS compatibility, <1V operation is necessary. This sets an upper bound for HZO film thickness of < 7 nm for $E_C = 1.5 \, MV/cm$. Non-resonant actuation of the devices would require a settling time of $\sim 2QT_0$. Assuming two operations per device, multiplication and addition, the computation speed (FLOPS per device) and power consumption can be estimated as $\frac{f_0}{Q}$ and $E\frac{f_0}{Q}$ respectively. Where E is the energy consumption per operation for individual device given by $\frac{1}{2}C_{HZO}.V_{in}^2$. For example, unimorph device of size 5 × 0.5 μm and thickness 28nm (~4×HZO thickness) with a Q=10, which can be realized by high-pressure gas damping, yields $f_0$ ~20MHz and computation speed of 2MFLOPS/device. Considering driving electrodes of size 1×0.3 μm with 7nm thick HZO between $V_{in}$ (0.1V) and GND electrodes, the energy consumption per operation can be calculated as ~500 aJ ($E = \frac{1}{2}C_{HZO}.V_{in}^2$). Combining both values result in 2 PFLOPS/watt computational performance. This computation speed with ultra-low energy consumption and zero leakage current will pave the path toward brain-level computation efficiency.

**Supplementary Material:**

See supplementary material for the detailed fabrication process and challenges faced.

**Acknowledgment:**

This work was performed in part at the Cornell NanoScale Facility, an NNCI member supported by NSF Grant NNCI-2025233. Funding was provided by the DARPA TUFEN program.

**References:**

[1] C. Mead, "Neuromorphic electronic systems," *Proc. IEEE*, vol. 78, no. 10, pp. 1629–1636, 1990.
[2] E. R. Kandel and J. H. Schwartz, *The Nervous System: Principles of Neural Science*, vol. 217, no. 4556. 1982.
[3] A. Keshavarzi, K. Ni, W. Van Den Hoek, S. Datta, and A. Raychowdhury, "FerroElectronics for Edge Intelligence," *IEEE Micro*, vol. 40, no. 6, pp. 33–48, Nov. 2020.
[4] W. Haensch, T. Gokmen, and R. Puri, "The Next Generation of Deep Learning Hardware: Analog Computing," *Proc. IEEE*, vol. 107, no. 1, pp. 108–122, 2019.
[5] M. Horowitz, "Computing ' s Energy Problem ( and what we can do about it )," pp. 10–14, 2014.
[6] M. Si, X. Lyu, and P. D. Ye, "Ferroelectric Polarization Switching of Hafnium Zirconium Oxide in a Ferroelectric/Dielectric Stack," *ACS Appl. Electron. Mater.*, vol. 1, no. 5, pp. 745–751, May 2019.
[7] M. Jerry *et al.*, "Ferroelectric FET analog synapse for acceleration of deep neural network training," in *2017 IEEE International Electron Devices Meeting (IEDM)*, 2017, pp. 6.2.1-6.2.4.
[8] S. S. Cheema *et al.*, "Enhanced ferroelectricity in ultrathin films grown directly on silicon," *Nature*, vol. 580, no. 7804, pp. 478–482, Apr. 2020.
[9] V. Gund *et al.*, "Multi-level Analog Programmable Graphene Resistive Memory with Fractional Channel Ferroelectric Switching in Hafnium Zirconium Oxide," in *IEEE Frequency Control Symposium (FCS)*, 2022.
[10] A. Chen, S. Datta, X. S. Hu, M. T. Niemier, T. Š. Rosing, and J. J. Yang, "A Survey on Architecture Advances Enabled by Emerging Beyond-CMOS Technologies," *IEEE Des. Test*, vol. 36, no. 3, pp. 46–68, 2019.
[11] A. Sebastian, M. Le Gallo, R. Khaddam-Aljameh, and E. Eleftheriou, "Memory devices and applications for in-memory computing," *Nat. Nanotechnol.*, vol. 15, no. 7, pp. 529–544, 2020.

[12] D. E. Nikonov and I. A. Young, "Benchmarking Delay and Energy of Neural Inference Circuits," *IEEE J. Explor. Solid-State Comput. Devices Circuits*, vol. 5, no. 2, pp. 75–84, 2019.

[13] A. Lal, J. Hoople, S. Ardanuç, and J. Kuo, "Computation Devices and Artificial Neurons Based on Nanoelectromechanical Systems," 2019.

[14] R. . Ballas, *Piezoelectric Multilayer Beam Bending Actuators*. 2007.

[15] V. Gund *et al.*, "Temperature-dependent Lowering of Coercive Field in 300 nm Sputtered Ferroelectric Al0.70Sc0.30N," in *IEEE International Symposium on Applications of Ferroelectric (ISAF)*, 2021, pp. 1–3.

[16] M. Ghatge, G. Walters, T. Nishida, and R. Tabrizian, "A Nano-Mechanical Resonator with 10nm Hafnium-Zirconium Oxide Ferroelectric Transducer," *Tech. Dig. - Int. Electron Devices Meet. IEDM*, vol. 2018-Decem, pp. 4.6.1-4.6.4, 2019.

[17] S. Starschich, T. Schenk, U. Schroeder, and U. Boettger, "Ferroelectric and piezoelectric properties of Hf 1-x Zr x O 2 and pure ZrO 2 films," *Appl. Phys. Lett.*, vol. 110, no. 18, p. 182905, May 2017.

[18] S. Riedel, P. Polakowski, and J. Müller, "A thermally robust and thickness independent ferroelectric phase in laminated hafnium zirconium oxide," *AIP Adv.*, vol. 6, no. 9, p. 095123, Sep. 2016.